# Novel Smart N95 Filtering Facepiece Respirator with Real-time Adaptive Fit Functionality and Wireless Humidity Monitoring for Enhanced Wearable Comfort

*Kangkyu Kwon[1,2†], Yoon Jae Lee[1,2†], Yeongju Jung[3], Ira Soltis[2,4], Chanyeong Choi[2,4], Yewon Na[2,4], Lissette Romero[2,5], Myung Chul Kim[2,4], Nathan Rodeheaver[2,4], Hodam Kim[2,4], Michael S. Lloyd[7], Ziqing Zhuang[8], William King[8], Susan Xu[8], Seung-Hwan Ko[4,9]\*, Jinwoo Lee[10]\*, Woon-Hong Yeo[1,11]\**

[1]School of Electrical and Computer Engineering, College of Engineering, Georgia Institute of Technology, Atlanta, GA, 30332, USA
[2]Center for Human-Centric Interfaces and Nanoengineering at Institute for Electronics and Nanotechnology, Georgia Institute of Technology, Atlanta, GA, 30332, USA
[3]Applied Nano and Thermal Science Lab, Department of Mechanical Engineering, Seoul National University, 1 Gwanak-ro, Gwanak-gu, Seoul, 08826, Korea
[4]George W. Woodruff School of Mechanical Engineering, Georgia Institute of Technology, Atlanta, GA, 30332 USA
[5]School of Industrial Design, Georgia Institute of Technology, Atlanta, GA 30332, USA
[6]Wallace H. Coulter Department of Biomedical Engineering, Georgia Institute of Technology, Atlanta, GA 30332, USA
[7]Division of Cardiology, Emory University School of Medicine, Atlanta, GA, 30322, USA
[8]National Institute for Occupational Safety and Health, National Personal Protective Technology Laboratory , Pittsburgh, PA, 15236, USA
[9]Institute of Advanced Machinery and Design (SNU-IAMD), Seoul National University, Gwanak-ro, Gwanak-gu, Seoul, 08826, Korea
[10]Department of Mechanical, Robotics, and Energy Engineering, Dongguk University, 30 Pildong-ro 1-gil, Jung-gu, Seoul, 04620, South Korea
[11]Wallace H. Coulter Department of Biomedical Engineering, Parker H. Petit Institute for Bioengineering and Biosciences, Neural Engineering Center, Institute for Materials, Institute for Robotics and Intelligent Machines, Georgia Institute of Technology, Atlanta, GA, 30332 USA

† K. Kwon and Y. J. Lee equally contributed to this work

\*Correspondence to Dr. Woon-Hong Yeo (whyeo@gatech.edu), Dr. Jinwoo Lee (jlee484@dgu.ac.kr), and Dr. Seung-Hwan Ko (maxko@snu.ac.kr)

**Abstract**

The widespread emergence of the COVID-19 pandemic has transformed our lifestyle, and facial respirators have become an essential part of daily life. Nevertheless, the current respirators possess several limitations such as poor respirator fit because they are incapable of covering diverse human facial sizes and shapes, potentially diminishing the effect of wearing respirators. In addition, the current facial respirators do not inform the user of the air quality within the smart facepiece respirator in case of continuous long-term use. Here, we demonstrate the novel smart N-95 filtering facepiece respirator that incorporates the humidity sensor and pressure sensory feedback-enabled self-fit adjusting functionality for the effective performance of the facial respirator to prevent the transmission of airborne pathogens. The laser-induced graphene (LIG) constitutes the humidity sensor, and the pressure sensor array based on the dielectric elastomeric sponge monitors the respirator contact on the user's face, providing the sensory information for a closed-loop feedback mechanism. As a result of the self-fit adjusting mode along with elastomeric lining, the fit factor is increased by 3.20 and 5 times at average and maximum respectively. We expect that the experimental proof-of-concept of this work will offer viable solutions to the current commercial respirators to address the limitations.

# 1. Introduction

The COVID-19 pandemic has emphasized the critical role of facial respirators as essential personal protective equipment in reducing the transmission of airborne pathogens(1, 2), and, facial respirators have become an integral part of daily life across the globe consequently(3). Despite their widespread adoption, the long-term use of conventional respirators, including disposable, cloth, and simple elastomeric designs, has exposed several limitations, associated with wearability and environmental monitoring capabilities. One of the most prominent issues with conventional respirators lies in the lack of proper fit and comfort of the commercial respirators that cannot accommodate the wide range of face sizes and shapes of the users(4, 5). Ill-fitting respirators frequently lead to gaps around the nose and mouth, compromising their effectiveness in filtering out harmful particles(6). Also, the individuals have varying facial features, such as small or large faces, and yet respirators are often produced in a 'one-size-fits-all approach'(7). Furthermore, the accumulation of moisture inside the respirator and fogging of eyewear is a common problem that can negatively impact the user experience, further highlighting the need for a more adaptable and personalized respirator. Another limitation of current respirator solutions is the absence of real-time environmental monitoring capabilities as conventional respirators are unable to provide users with valuable information about the surrounding air quality, such as humidity that could be an essential indicator for ensuring optimal respirator performance and user's wear comfort.

Efforts in current research and product development have begun addressing the limitations of commercially available facial respirators. Notably, some recent studies have investigated the incorporation of sensors and electronics into face respirators to monitor respiratory activity(3, 8-13), respirator fit (6, 14, 15), and other vital signs(16-23). Nevertheless, these advancements frequently involve complicated and bulky electronic components that can aggravate their overall wear comfort. Besides, the previous solutions lack real-time feedback mechanisms, so the users do not receive immediate updates regarding the status of the facepiece respirator fit or breathing air quality.

Given the inherent limitations in the designs of previous respirators, there has been an increasing demand for the integration of wireless real-time wearable technologies into respirator designs. Parallel to this, there is a growing interest in the development of smart respirators equipped with environmental sensors such as carbon dioxide(21, 24, 25) and humidity(13, 22, 26) for monitoring the air quality within the respirator. Yet, these sensors often involve rigid design and do not provide wireless wearable monitoring and sensory feedback that enable the respirator to self-adjust its fit on the user's face based on sensor readings. In this regard, despite the significant advance in respirator technologies, the absence of adaptive respirator designs and the lack of environmental monitoring represent key limitations of current approaches.

Here, we demonstrate a novel smart N-95 filtering facepiece respirator design that integrates LIG sensors for real-time humidity monitoring along with a unique adaptive fit mechanism using pressure sensors and actuators to enhance wearability and ensure optimal respirator performance. The LIG sensor offers high sensitivity and selectivity for humidity measurements while retaining lightweight and flexible properties that allow continuous monitoring of the user's breathing environment as an indicator of respirator performance and overall air quality. The adaptive fit mechanism, which utilizes eight pressure electrodes to detect the user's facial contours and two motors to adjust the respirator, ensures a secure and comfortable seal around the face, mitigating common issues such as respirator shifting, fogging eyewear, and skin irritation. The smart respirator design proposed in this work offers a more comprehensive solution with the potential to significantly improve the overall experience of wearing personal protective equipment. Table 1 epitomizes the unique aspects of this work by drawing a comparison with the existing respirator research. The shaded areas in the table demonstrate the desired capabilities of a smart respirator that offers improved comfort and air quality monitoring.

We believe that the facepiece respirator system investigated in this work will serve as the exclusive solution that matches all specifications for a user-driven and efficient protective respirator.

## 2. Result and discussion

### 2.1 General overview of the smart N-95 filtering facepiece respirator system

**Figure 1a** presents the graphical illustration that describes each component of the smart facepiece respirator and delivers the general concept of this work. As depicted in the illustration, the basic framework of the smart facepiece respirator consists of the pressure sensor array, a humidity sensor based on LIG, flexible circuits, and a fit-adjusting actuator, all of which are integrated into the commercially available N-95 respirator. The flexible circuit based on the polyimide (PI) substrate collects the facial fit and humidity data within the smart respirator in real time and wirelessly transmits the acquired data to a portable device such as a smartphone. If the facepiece respirator is not worn properly, the smartphone detects the poor fit of the respirator based on the pressure sensor array and autonomously adjusts the respirator fit by operating actuators through wireless Bluetooth communication. In addition to the self-adaptive fit functionality, the smartphone simultaneously updates the moisture level and informs the user about the real-time air quality within the respirator. Lastly, all of the acquired humidity and pressure sensing data is stored in the cloud storage for future reference, of which the personal data is only accessible by the user or authorized staff in charge.

The flowchart in **Figure 1b** describes how each module constructs the sensory feedback and how the smart respirator operates as a whole. LIG sensor monitors the moisture level while the pressure sensor electrodes keep track of the respirator contact level of the eight points on the user's face. The flexible circuit collects the acquired humidity and pressure data and wirelessly transfers them to the portable device, where the data is simultaneously processed for the next step. The processed data is saved in the cloud storage and also wirelessly transferred to the fit-adjusting actuator modules for sensory feedback control. In case of a poor respirator fit, servomotor actuation takes place to translate force to the smart respirator via straps such that the respirator autonomously adjusts the respirator fit.

**2.2 Smart N-95 filtering facepiece respirator component overview and pressure sensor characterization**

**Figure 2a** presents the graphical representation of the smart facepiece respirator and its constituting components such as a customized magnet chargeable battery, flexible pressure sensor circuit, and flexible electrode that interconnects the pressure sensor array. The flexible circuit includes microchips that enable the real-time data collection of pressure data and wireless data transfer to the portable device. As illustrated in **Figure 2b**, the pressure sensor, which consists of the two PI/Cu electrodes with the dielectric elastomeric sponge in between, operates by measuring capacitance increase upon applied pressure as the Ecoflex sponge elastically deforms in the axial direction. We fabricated the PI/Cu electrode by depositing Cu on the PI film and then by laser-cutting it along the designed pattern. As for the Ecoflex sponge, in a 1:1 mixture of sugar and pre-cured Ecoflex, the sugar remains undissolved and thus maintains its crystalline structure throughout the curing process in a vacuum. After curing, the elastomer-sugar composite was then immersed in a hot water bath to dissolve sugar, resulting in a compressible, spongy elastomer foam. Lastly, the elastomer foam is cut into the desired design and inserted in between the PI/Cu electrodes as a dielectric material for capacitance measuring. **Figure 2c** shows the photographs of the PI/Cu electrode and Ecoflex sponge that constitute the pressure sensor array. The Ecoflex sponge is placed on top of the PI/Cu electrode, and the counterpart PI/Cu electrode is then folded over the Ecoflex sponge such that it is sandwiched between two electrodes. Besides, the sensor array design also includes the connection pad to fit into the ribbon cable for facile integration into the circuit (Figure S1). Lastly, the Silbione and Ecoflex encapsulation on top of the pressure sensor array enhances intimate contact between the smart respirator and the user's skin surface.

To determine the optimal thickness of the Ecoflex sponge and to further verify the performance of pressure sensing, we measured the capacitance of pressure sensors with different thicknesses (1.0 mm, 1.5 mm, 2.5 mm, and 3.0 mm) while varying the applied force. The thickness comparison result suggests that the elastomeric sponge with a thickness of 1.0 mm demonstrates the greatest capacitive response when a given range of force was applied to the sensor (See **Figure S2** for the experimental setup). The aforementioned result follows our theoretical expectations on the sponge thickness because thinner samples would deform much easier than thicker ones if we assume that the mechanical modulus of the material is uniform. In this regard, we selected a sponge thickness of 1.0 mm and characterized a pressure sensor based on it. **Figure 2e** exhibits the normalized capacitance of the pressure sensor as it undergoes repeated application of pressure over 5000 cycles. As can be observed from the experimental results, the pressure sensor demonstrates substantially consistent normalized capacitance change against the fatigue stress throughout the entire set of cycles, highlighting the reliability of the pressure sensor array to monitor the respirator fit on the user.

**2.3 Fabrication and characterization of LIG sensor for humidity monitoring**

**Figure 3a** explains the simple fabrication method to make a graphene sensor by irradiating laser on the PI substrate. The graphene sensor in this work also includes the heater component on the other side of the sensor because high temperature facilitates the desorption of gaseous molecules from the graphene sensor(27, 28). Thus, after the graphene sensor is inscribed on PI by laser irradiation, laser is irradiated on the other side of the sensor to make a graphene heater as depicted in the figure. **Figure 3b** shows the photographs of the sensor and heater, which respectively correspond to the top and bottom sides of the humidity sensor. The same laser irradiation conditions were employed to fabricate the sensor and heater, but their respective designs differ from one another. For instance, for the humidity sensor design, the two contact electrodes were connected by a thin, elongated electrode with a width of 200 μm (as in the optical microscope image), which enables sensitive measurement of humidity and allows the electrode to capture even a small difference of moisture level. On the other hand, the heater design takes a form of a rectangular block because it requires uniform heating of the entire PI film such that it accelerates the desorption of water molecules when the smart respirator is not in use. The scanning electron microscope (SEM) image in **Figure 3b** exhibits the extremely porous structure of LIG that serves to enhance the sensitivity of the humidity sensor as the porous structure provides a high surface area for the water molecules to adhere to.

**Figure 3c** shows the Raman spectrum that offers valuable information to determine the existence of graphene during laser irradiation on PI. The Raman spectrum of graphene includes unique peaks such as the D band (around 1340 cm$^{-1}$), G band (around 1585 cm$^{-1}$), and 2D band (around 2696 cm$^{-1}$) as labeled in the figure. Normally, when characterizing the graphene with Raman spectrum, the D band spectrum signifies the existence of either sp$^2$ hybridization C bonds or defects, whereas the G band corresponds to the 1$^{st}$-order phonons, and the 2D band represents the 2$^{nd}$-order boundary phonons.(29) The ratio of $I_{2D}/I_G$ of 0.620 indicates the formation of single-layer graphene as the value is considerably lower than 2. In addition, compared to the ideal graphene that has the 2D band at 2685 cm$^{-1}$, the 2D band in this work shows a blueshift of approximately 11 cm$^{-1}$, which appears to be attributed to compressive strain due to the rapid heating and subsequent cooling process during the laser irradiation. (30, 31)

To examine the performance of the humidity sensor based on LIG, we measured the normalized resistance of the sensor as we increased the humidity level over time. **Figure 3d** exhibits the corresponding normalized resistance as the moisture concentration level is step-wisely increased. For graphene as a humidity sensor, water molecules act as electron donors, so the electron transfer from water molecules to graphene serves to increase the bandgap of graphene, leading to its resistance increase when the humidity level rises.(32) Thus, as the humidity level increased in the unit of percentage from 0% to 80% with an interval of 20%, the normalized resistance of the LIG sensor rose accordingly with different humidity levels. In addition to testing the humidity sensor performance, we also characterized the performance of the LIG heater as in **Figure 3e**, in which the electrical voltage was increased discretely. As a result, the temperature of the graphene heater increased in a proportional manner to the applied voltage as depicted by the thermal imaging snapshots, and the heater approached approximately 66°C at 6 V or 163 mW and all the way up to 125°C at 10 V or 434 mW, implying that the graphene heater can produce a highly controllable heating profile with low power consumption to accelerate desorption process.

**2.4 Operation logic and component integration into smart N-95 filtering facepiece respirator system**

As illustrated in **Figure 4a**, the firmware design of the smart facepiece respirator system incorporates several functional layers to ensure smooth operation for monitoring the humidity sensor, pressure sensor array, and autonomous fit adjustment capability. The functional back-end layers of the firmware encompass the processing of raw sensor data, execution of device functions and feedback control, as well as power management for optimal energy consumption. To streamline communication tasks, a low-power system-on-chip (SoC) microcontroller unit (MCU) serves as the main processor unit with wireless Bluetooth low energy and multiple communication layers. The sensor data processing layer is responsible for processing raw data obtained from embedded sensors, while the control layer governs the execution of various functions based on sensor data and received commands.

Demonstrating pressure sensing in wireless real-time smart respirators involves implementing an integrated FDC1004 sensor IC component, interfaced with an MCU and the two-wire interface (TWI). The pressure sensor system incorporates an onboard adaptive sensing solution to enable automatic calibration of pressure measurements. This calibration process aims to minimize the effects of environmental fluctuations and variations in respirator conditions on the capacitance measurement system. To enhance system accuracy, a closed-loop feedback mechanism is designed to enable active communication from the pressure sensor device to the Android application. Upon establishing a connection between the pressure sensors and the portable device, the system automatically performs calibration for the capacitance of each individual respirator. This calibration process utilizes the FDC1004 firmware protocol to adjust the off capacitance and capacitive gain. Besides, to facilitate LIG resistance sensing, the system incorporates an ADS1292 sensor, which serves as a low-power analog front-end (AFE) sensor through a serial peripheral interface (SPI). Additionally, a Wheatstone bridge circuit is utilized to enable the MCU to accurately measure electrical resistance values. Lastly, servomotor drivers are programmed to govern the motor operation by transmitting digital signals through the general-purpose input/output (GPIO) pins of the MCU.

As illustrated in the figure, integration with a cloud service allows the data to be stored remotely in real time and enables large-scale data analysis, machine learning applications for statistical analysis, and the ability to provide personalized feedback to users based on historical data. Moreover, healthcare professionals or relevant authorities could access and analyze this data remotely, enhancing monitoring and intervention strategies. Anonymized cloud-stored data is used to gain insights into broader trends and patterns, thus contributing to improving the overall design and functionality of smart respirators and public health policies regarding respirator usage.

The proposed concept requires two different flexible printed circuit board (fPCB) designs with wireless communication capability: one for the pressure sensor and the other for the fit-actuator/LIG humidity sensor. **Figure 4b** presents the graphical illustration of the fPCB for real-time data collection from the pressure sensor array and wireless data transfer to the designated portable device. The fPCB consists of an MPU and Bluetooth antenna for wireless data transfer. Here, the MPU receives the pressure sensor data from the two capacitive measurement ICs that are connected to two flexible flat cables (FFC) with 5 pins, each of which monitors a right and left pressure array. The FFC 5 pins form a stable electrical connection with the ribbon cable, which can be easily integrated with the connection pad in the PI/Cu electrode. Similarly, the fPCB for the fit-actuator/LIG sensor includes identical MPU and antenna components for wireless data transfer, but it consists of two modules for actuating servomotors and humidity sensing as can be observed in **Figure 4c**. The resistance measuring IC monitors the real-time resistance of the LIG electrodes that are electrically connected to the circuit with the LIG sensing contact pads via ribbon cables. **Figure S3** shows the wireless circuits for humidity sensing/fit actuator and 8-channel pressure sensing that are encapsulated with the white-dyed Ecoflex. Besides, the heating contact pads are used to supply electrical voltage for heating the electrode for gas desorption. Finally, **Figure 4d** exhibits the exterior view of the integrated smart respirator that is comprised of the fit-actuators and two fPCBs for pressure sensing and fit-actuator/LIG humidity sensor. **Figure S4** includes the interior snapshot of the smart respirator that consists of the LIG humidity sensor at the center and the pressure sensor array at the edge for respirator-fit monitoring in real time. The miniaturized incision is made on the respirator to connect the LIG sensor/heater electrode to the circuit that is attached to the exterior of the smart respirator. The customized magnet rechargeable battery developed in this work allows easy recharging of circuits with a magnetic cord as in **Figure S5.**

**Figure 4e** presents the flow chart that can be largely subdivided into self-fit adjusting and humidity monitoring modes. For self-fit adjusting mode, the respirator system monitors the real-time respirator contact status of the 8 points on the respirator perimeter with the user's facial skin. In the case of the loose-fit scenario, the servomotors continue actuating in real-time such that they fasten the strap until the contact status of 8 points approaches the proper fit. Finally, when the desirable fit has been attained either through servomotor actuation or deactuation, the smart respirator switches to the humidity monitoring mode. Here, if the humidity level within the respirator is too high, the servomotors loosen the straps and inform the user to take the respirator off. Then, the graphene heater on the other side of the humidity sensor starts heating up the sensor such that the water molecules become desorbed from the LIG electrode. After a period of stabilization, the water molecules on the LIG become completely desorbed by continued heating, and the smart respirator can be used again.

**Figure 4f** delineates the captured screen of the Android application that is capable of monitoring 8-point contact data and humidity level within the respirator in real time. With both onboard circuits that are wireless connected to the portable device, the smart respirator starts calibrating the initial capacitance values of each pressure sensor such that the smart respirator can keep track of the real-time change in capacitance when the user wears it. The color of the square blocks around the graphical representation of the respirator in **Figure 4f** indicates the degree of respirator fit. A color-coded fit alert system further enhances the user experience by offering real-time feedback, thereby aiding users in making the necessary adjustments to achieve a proper respirator fit. The green square block in the graphic user interface (GUI) signifies that the pressure sensor is calibrated and properly fit, while the dark gray block represents the over-fit. Likewise, the red and yellow blocks suggest a poor fit and a moderately poor fit respectively. When the user wears the respirator, the Android application informs the user whether the calibration is completed, and it also displays the real-time humidity within the respirator as in the top corner of the figure. Along with the auto-fit functionality, the application also allows manual fastening or loosening of the straps by controlling the left or right servomotors just in case the user intends to control the contact fit of the respirator in a desired manner. **Figure S6** includes a more detailed explanation of the GUI for the Android application.

**2.5 In vivo performance evaluation of the integrated smart respirator platform on human subjects**

**Figure 5a** delivers the front and side views of the integrated smart respirator worn by the human subject along with the exterior of the smart respirator which includes two wireless communication circuits and self-fit adjusting actuators. **Figure 5b** and Video S1 corroborate the validity of the self-fit adjusting mode presented in this work. In Video S1, the screen of the tablet PC exhibits that each of the 8-point pressure sensors turns from blue to red one by one, suggesting that all the pressure sensor units are wirelessly connected to the tablet PC. When the human subject wears the respirator, some of the pressure sensor units are in yellow and red, indicating a poor respirator fit. Yet, in the presence of the self-fit adjusting mode, the servomotors fasten the strap to attain a proper fit, and all the pressure sensors in the screen of the tablet PC turn green. Even if the human subject in the video wears a beard that usually interferes with the proper respirator fit, the self-fit adjusting mode ensures intimate contact with the user's skin, thereby effectively blocking the potential pathogens from penetrating the respirator through the interfacial gap. Also, if the human subject tries to move the respirator as in the video clip, the self-fit adjusting mode activates again and autonomously adjusts the respirator fit. To examine whether the self-fit adjusting mode of the smart respirator works for human subjects with different facial shapes and sizes, we recruited female and male subjects and have them operate the self-fit adjusting mode just as in Figure 5b and Video S1. Video S2 validates that self-fit adjusting mode functions just as well regardless of gender, face shape, and size, verifying the 'one-size-fits-all' approach truly works for the respirator that we developed in this study. Besides, in case of a respirator over-fit, or if the strap tension of the respirator is too high, the user can simply relocate the respirator, and the smart respirator automatically finds a more suitable respirator fit based on sensory feedback control. In Video S3, after the user relocated the respirator, the dark gray block (corresponds to over-fit) turns green, indicating that the respirator fit is converted into the proper fit after relocation.

In addition, we conducted the respirator fit test of the smart respirator developed in this work using the commercially available product (PortaCount™ Respirator Fit Tester 8048) that evaluates the respirator fit in real-time by comparing the concentration of microscopic particles in the ambient atmosphere and the concentration of the particles that leak into the facial respirator (See **Figure S7** for the equipment and its GUI). Thus, the fit factor is defined as the ratio of the particle concentration in the ambient atmosphere and the particle concentration that leak out. For instance, a fit factor of 10 suggests that the air inside the respirator is 10 times cleaner than the air outside the respirator.(33) **Figure 5c** compares the fit factor between the normal N-95 respirator and the smart respirator developed in this work during normal breathing, deep breathing, or performing specific activities that the user was asked to perform. The result draws a dramatic difference between the normal respirator and the smart respirator for all activities the user performed. For the normal breathing condition, the smart respirator in this work scored the fit factor that was 2.85 folds higher than the normal respirator. The fit factor gap widened especially when the user was asked to move the head side to side and up and down: the smart respirator generated a fit factor that was 3.25 times and 4 times higher than the normal respirator for head-side-to-side and head-up-and-down conditions respectively. The difference in the fit factors mainly arises from two factors: (i) the self-fit adjustable mode and (ii) the Silbione-Ecoflex encapsulation lining on the periphery of the smart respirator that serves to enhance conformal contact with the human skin as it has been previously reported that biogel such as elastomeric composite usually requires less debonding energy from the animal skin.(34)

**Figure 5d** shows the change in the normalized resistance of the LIG humidity sensor within the respirator when the human subject wears the smart respirator for three hours, takes it off, and leaves it in the ambient atmosphere. As expected, shortly after the human subject wears the respirator and continues breathing, the humidity level escalates sharply and then exhibits a gentle increase at a moderate slope. The gentler slope at the later stage appears to originate from the fact that water vapor is trapped inside the smart respirator since water molecules can not pass through the N-95 textile, thereby adding thermal stress and causing discomfort due to poor air quality. After wearing the smart respirator for 3 hours, the human subject was asked to take the respirator off, and the graphene heater embedded in the sensor starts to heat up at 60°C, accelerating the desorption process of water molecules. As a result, normalized resistance experiences a dramatic decline toward the baseline, and it remains constant when it was left in the ambient atmosphere. In this regard, we corroborate the practical usage of the smart N-95 filtering facepiece respirator in self-adjusting the respirator fit and enhancing the wear comfort by moisture level. Based on the current study, the future study will include the incorporation of multiple gas sensors into the respirator system such that they can be utilized to quantify the air quality within the respirator in more detail and also work in conjunction with the self-adjusting mode of the current system.

## 3. Conclusion

In this work, we develop a novel smart N-95 filtering facepiece respirator platform that integrates the LIG-based humidity sensors and pressure sensor array to self-adjust the respirator while regulating the air quality within the respirator for optimal respirator performance. The pressure sensors based on the dielectric elastomer sponge form an array at the periphery of the respirator for monitoring the real-time respirator fit by measuring the capacitance of the pressure sensing electrodes. Along with the pressure sensing array, the LIG-based humidity sensor detects the rising humidity level within the respirator, and the embedded graphene heater can accelerate the water desorption rate such that the humidity level can be stabilized after use. The onboard flexible circuits collect real-time pressure and humidity data and wirelessly transmit them to the portable device, which processes the data into meaningful information to be used for self-fit adjusting and humidity monitoring modes. The real-time demonstration of the self-fit adjusting mode on the human subjects highlights that such a mode operates just as well regardless of gender, facial shape, size, and the presence of a beard. Besides, the fit factor comparison between the smart respirator and the commercially available respirator validates the practical effectiveness of the smart respirator since the fit factor of the smart respirator was 4 times higher at maximum when compared to the commercial respirator, owing to the self-fit adjusting mode and elastomeric lining at the periphery of the smart respirator. Furthermore, the humidity monitoring test while wearing a smart respirator substantiates that the embedded sensor is capable of tracking the humidity level in real time with high reliability. Lastly, considering the practical usage of the smart respirator in real-life situations, the smart respirator is anticipated to address significant challenges faced by frontline healthcare workers, particularly those related to maintaining a secure face seal while wearing a respirator. The in vivo validation with human subjects in comparison with the commercially available respirator further highlights the effectiveness of the smart respirator. In this regard, we expect that the smart respirator with novel functionalities proposed in this work will not only address the current limitations and challenges of the current respirators but also offer valuable insights for future bioelectronics.

## 4. Materials and Methods

### 4.1 Pressure sensor fabrication and characterization

The fabrication of the capacitive pressure sensor for the smart respirator involves two primary processes: 1) sensor array electrode fabrication, which comprises the production of the electrode and dielectric layer, and 2) circuit fabrication and sensor encapsulation.

For the fabrication of the electrode layer fabrication, polyimide (PI) and the copper (Cu) layer, designated CAD design, were precisely shaped through a laser cutting process. The interconnector parts were subsequently encapsulated on both sides with polyimide film, while the sensor electrode parts received polyimide encapsulation on one side only. To encapsulate the pressure sensor array, the two sensing electrodes were folded together with an Ecoflex sponge positioned in between as a dielectric layer of the pressure sensor. This assembly was combined with the respirator using elastomer (Ecoflex 00-30, Smooth-On) and Silbione (A-4717, Factor II Inc.).

For the fabrication of circuits and sensor encapsulation, an fPCB was utilized. All electronic components were mounted onto the fPCB using a reflow soldering process. Unnecessary regions of the PCB were then removed via laser cutting to augment the mechanical flexibility of the circuit. Power supply and management were addressed through a lithium polymer battery assembly, which was furnished with a slide switch and a circular magnetic recharging port. A low-modulus elastomer (Ecoflex Gel, Smooth-On) was positioned beneath the integrated circuit as a means to lessen strain. The complete electronic system was encapsulated and soft-packaged using an additional elastomer (Ecoflex 00-30, Smooth-On), leaving only the switch and the charging port exposed.

For measuring mechanical properties, the experimental setup included a digital force gauge (M5-5, Mark-10) and a motorized test stand (ESM303, Mark-10). Additionally, an LCR meter (Model 891, BK

Precision) was utilized to measure electrical capacitance. For the fatigue test, the flat tip of the force gauge was moved vertically with a speed of 20 mm/min for 5000 cycles.

**4.2 Laser-induced graphene sensor**

We utilized a 532 nm Nd:YAG continuous-wave laser (Lighthouse Photonics Inc.) to generate graphene on a PI film (18-2F, CS Hyde). The laser beam traversed an optical system that incorporated a galvanometer scanner and a telecentric lens-equipped galvano-mirror scanner system. Control over laser power, scan speed, and the patterning of pre-designed CAD models was achieved via computer-aided patterning software that was integrated with the galvano-mirror scanner. The PI film was transformed into laser-induced graphene (LIG) by irradiating the laser beam (Power-150mW, Scan rate 150mm/s) using this patterning system. The fabrication of the circuit and sensor integration process followed the same steps as those used in pressure sensor production.

For the analysis of the surface and cross-section morphology, we employed a Field Emission scanning electron microscope (FE-SEM, SUORA 55VP; ZEISS) and an optical microscope (OLYMPUS BX51). The molecular fingerprint was investigated using Raman spectroscopy, performed with the Raman microscope (inVia Raman microscope; Renishaw), exploring the interaction between light and molecules. For the Joule-heating experiment, we employed a DC Power supply (KEITHLEY; 2231A-30-3) and IR camera (FLIR A645) to measure temperature of the LIG heater on the other side of the sensor electrode by applying a voltage of 2 V to 10 V with the voltage interval of 2 V.

**4.3 Humidity sensing test**

A custom-built gas chamber was employed to regulate the flow of a specific volume of gas to the sensor for the humidity sensing test. The vapor gas flow rate was precisely controlled by two Mass Flow Controllers (MFCs) connected to humid air and dry air gas cylinders. The humid air was mixed with oxygen, and the dry air was mixed with nitrogen, with water present in the barrels. To verify the reproducibility of the LIG sensor in humidity detection, the vapor gas valve was meticulously sealed and set to 0% humidity prior to each change in humidity level. We explored humidity levels ranging from 20% to 100%, increasing in increments of 20%. A source meter (Model 2400; Keithley Instruments) was interfaced with the LIG sensors, and a computer, operating custom-designed software, was utilized to continuously monitor variations in sensor resistance.

**4.4 Firmware and layer structure for data acquisition system.**

The firmware design for the smart respirator system incorporates multiple functional layers, ensuring seamless operation for humidity sensor monitoring, pressure sensor array, and autonomous fit adjustment functionality. Functional back-end layers on firmware include raw sensor data processing, control execution for device functions and feedback control, and power management to optimize energy consumption. With a low-power SoC microcontroller unit as the main processor unit (NRF52832-Nordic), the wireless BLE (Bluetooth Low Energy) protocol and multiple communication layers were developed to streamline communication tasks. The sensor data processing layer handles and processes raw data from embedded sensors, and the control layer control executions by computing various functions based on sensor data and received commands, triggering responses.

**4.5 Wireless real-time smart respirator operation and autonomous fit adjustment.**

Demonstrating pressure sensing in wireless real-time smart respirator involves implementing an integrated FDC1004 sensor IC component, interfaced with a microcontroller (MCU), using the two-wire interface (TWI) protocol. The sensor measures variations in capacitance and subsequently converts these changes into digital signals, a format that can be interpretable and utilized through device firmware. For LIG sensing, serial peripheral interface (SPI) interfaced ADS1292, a low-power analog front-end (AFE) sensor, and the Wheatstone bridge circuit allows the MCU to measure precise electrical resistance values. Moreover, motor drivers (PDRV8210DRLR) are programmed to control the operation of the motor by sending digital signals via the MCU GPIOs.

To enhance system accuracy, sensor detection sensitivity is optimized through calibration and adaptive sensing. Upon launching the Android application, the device automatically calibrates the off capacitance and capacitive gain of the individual respirator by utilizing the FDC1004 firmware protocol. Capacitive gain calibration is additionally computed for normalized capacitance measurement that updates the gain coefficient in a one-time programmable (OTP) memory.

A user-friendly Android application has been developed to include key functionalities that consist of real-time pressure detection, wireless data transmission, and recording, intuitive visualization with pressure graphs with alert arrays, and adaptive capacitive sensing with fitting alerts. The application facilitates on-demand monitoring of pressure data and provides real-time alerts by communicating with the user interface (UI) layer of firmware for interaction, and the power management layer for optimizing energy consumption. These layers work together to provide a seamless user experience with the NRF52832 MCU's potent processing capabilities and energy efficiency.

A color-coded fit alert system, employing red, yellow, and green indicators, offers users an intuitive method to gauge fit accuracy. A red indicator signifies a poor fit requiring adjustment, a yellow indicator suggests an acceptable fit with room for improvement, a green indicator confirms a good fit and a dark gray for over fit. The color-coded fit alert system further provides real-time feedback, assisting users in making necessary adjustments for proper respirator fit.

Leveraging various the human subject dataset, the incorporated system support real-time pressure monitoring and feature detections. Furthermore, the application empowers users to rapidly assess the quality of the face seal and make necessary adjustments by altering the elastomer seal or tightening the bands for an optimal fit.

For the respirator fitting performance on the Auto-fit mode, the fit factor of the normal respirator and our smart respirator has been measured using PortaCount (PortaCount Pro+ Respirator Fit Tester 8038). We have followed the respirator testing protocol of "OSHA 29CFR1910.134", which measures the fit factor of a respirator for 4 different circumstances/actions.

### 4.6 Data cloud computing interface and software application.

Pressure and fit testing results, as well as humidity monitoring integrated into a cloud service, significantly enhance the usability and functionality of a smart respirator system. The user's data, including pressure, fit test results, and humidity levels, is automatically saved to a cloud server. This not only ensures the data's safety and accessibility, even if the local mobile device is damaged or lost but also enables users to access their data from multiple devices. This would also provide an easy way for users to track their respirator usage over time and monitor changes in their fit test results and humidity levels.

By integrating with a cloud service, the data is collected and stored remotely in real time. This allows for large-scale data analysis, machine learning applications for statistical analysis, and the ability to provide personalized feedback to users based on historical data. Moreover, healthcare professionals or relevant authorities could access and analyze this data remotely, enhancing monitoring and intervention strategies. Anonymized cloud-stored data is used to gain insights into broader trends and patterns. This data could contribute to improving the overall design and functionality of the smart respirators and inform research and public health policies regarding respirator usage.

**Table 1**: Comprehensive comparison between this work and previously reported works on smart respirator and proposed novel functionalities

| References | Integrated system* | Adaptive fit system | | Sensor feedback | | | Gas sensing capability | |
|---|---|---|---|---|---|---|---|---|
| | | Real-time automatic fitting | Cloud Storage | Sensor type | Material | Wireless | Material | Wireless |
| This work | Yes | O | O | Pressure | Cu / Sugar + Ecoflex | O | LIG | O |
| (24) | X | X | X | - | - | X | $La_2O_2S$ | O |
| (12) | X | X | O | Pressure | Au / Teflon AF | O | - | X |
| (35) | X | X | X | X | - | X | WCNT | X |
| (36) | X | X | X | X | - | X | RGO/ $SnO_2$ | O |
| (37) | X | X | X | Temperature | Commercial | X | ZIF-8 / PAN | O |
| (26) | X | X | X | X | - | X | Porous Silicon | O |
| (14) | X | O | X | X | - | X | - | X |
| (8) | X | X | X | Pressure | CNT/PDMS | O | - | X |
| (38) | X | X | X | Pressure | MTP | O | - | X |
| (15) | X | X | O | Strain | Ag / PI | O | - | X |
| | | | | Temperature | Ag / PI | O | | |
| (20) | X | X | X | Ultrasonic | Commercial | O | - | X |
| | | | | Temperature | Commercial | O | | |
| (13) | X | X | X | X | - | X | PP / Si | O |
| (9) | X | X | X | Respiratory | RS-TENG (FEP / Al) | X | - | X |
| (16) | X | X | X | Harmonic | AgNW | O | - | X |
| (11) | X | X | X | Strain | GNP/PCL | O | - | X |

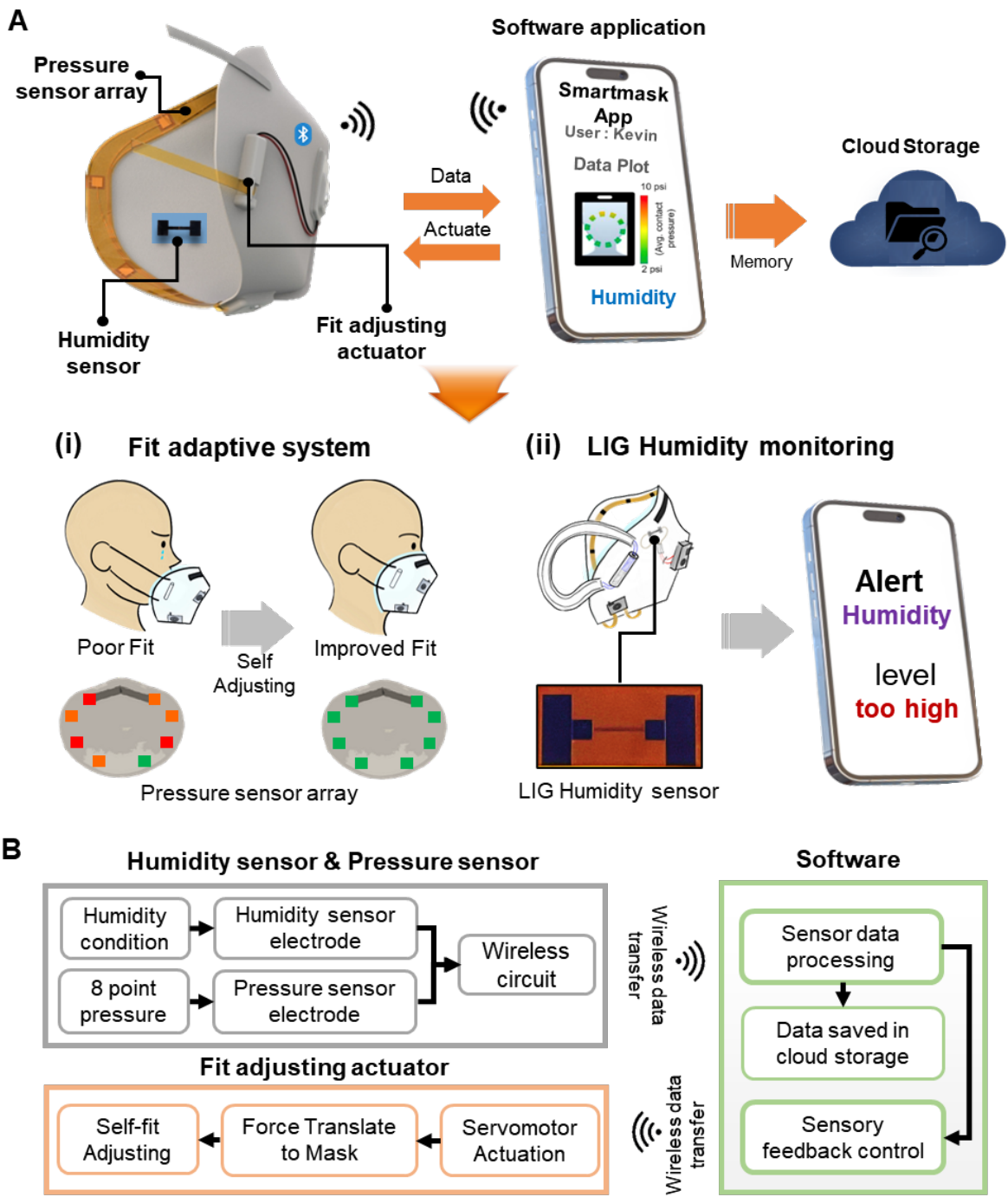

**Figure 1**

General overview of the smart respirator and its operating mechanism. A) Conceptual illustration of the wireless smart respirator in this work that can monitor humidity within the respirator and self-adjust the respirator fit based on sensory feedback control. The embedded wireless circuits transfer the collected data to the portable device where the data is both processed for sensory feedback and stored in the cloud storage for future reference and personal access. B) Flow chart that describes the operative mechanism of the smart respirator.

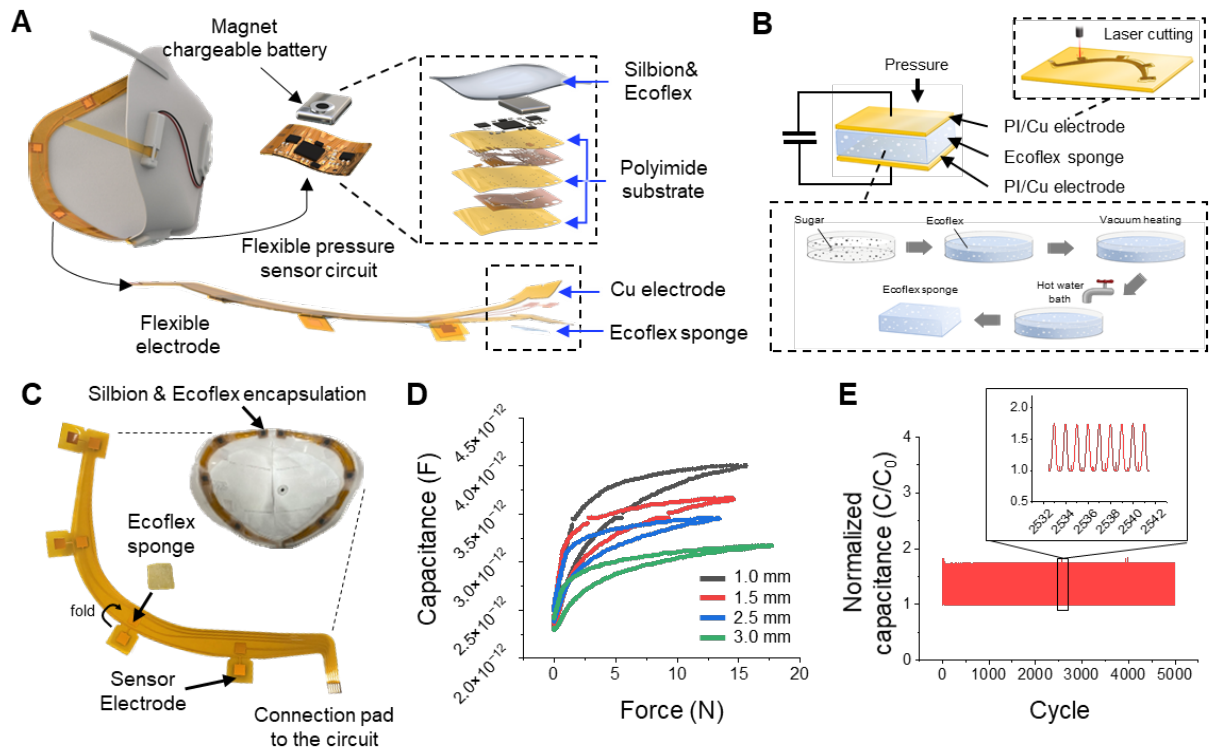

**Figure 2**

Pressure sensor array for sensory feedback-enabled self-respirator fit adjustment. A) Exploded view of the smart respirator that is comprised of the flexible pressure sensor circuit, flexible electrode, and magnet chargeable battery. B) Fabrication process to make a single unit of pressure sensor. C) Snapshot of the pressure sensor array along with the interior view of the smart respirator. D) Capacitance change in relation to force for various dielectric elastomer sponge thicknesses. E) Cyclic test of normalized capacitance when constant force is repeatedly applied to the sensor.

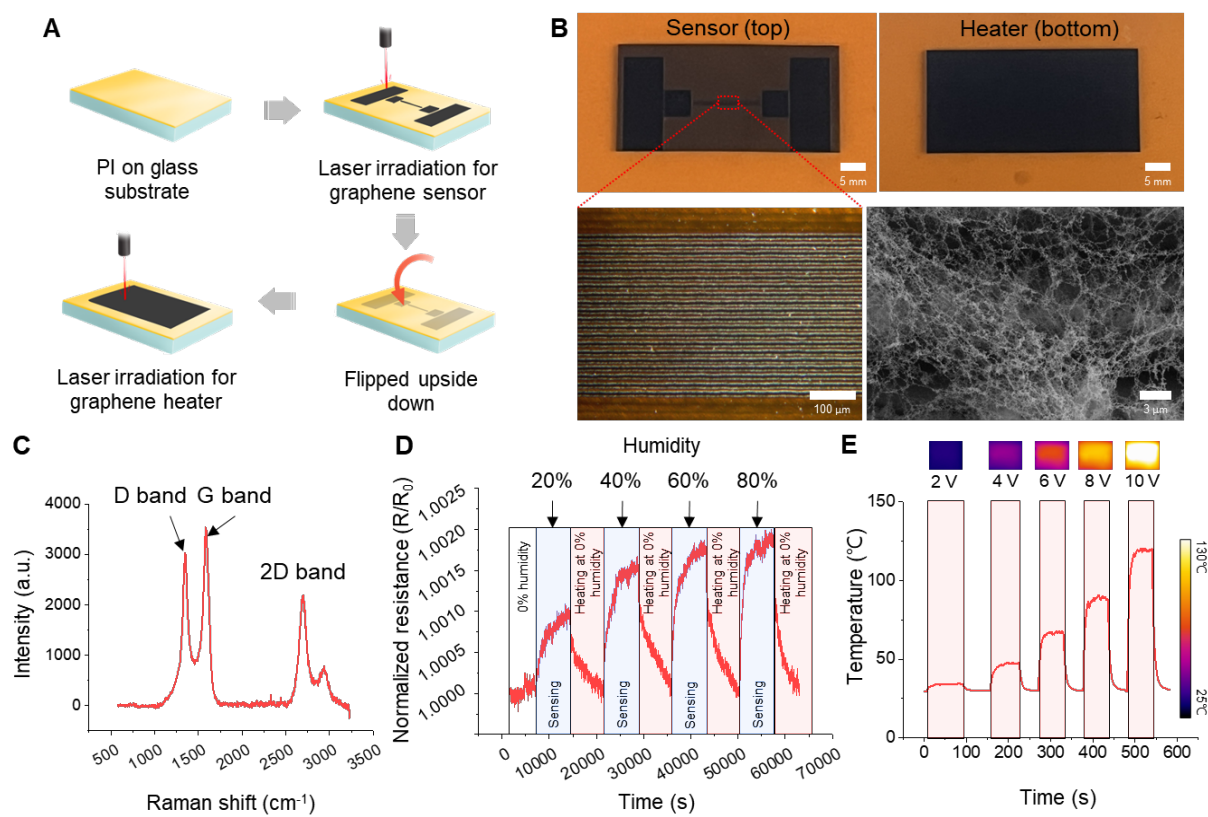

**Figure 3**

LIG characterization as a graphene humidity sensor. A) Fabrication process for the LIG humidity sensor and heater. B) optical microscope of LIG humidity sensor and heater along with the SEM image of the LIG electrode. C) Raman spectroscopy of the LIG electrode. D) Normalized resistance of the LIG humidity sensor as the humidity level increases from 0% to 80% with an increasing interval of 20%. For each interval, the sensor was heated at 60°C and 0% of humidity to facilitate the desorption of the water molecules from the LIG electrode. E) Temperature profile of the LIG heater when the electrical voltage from 2 V to 10 V is applied to the heater.

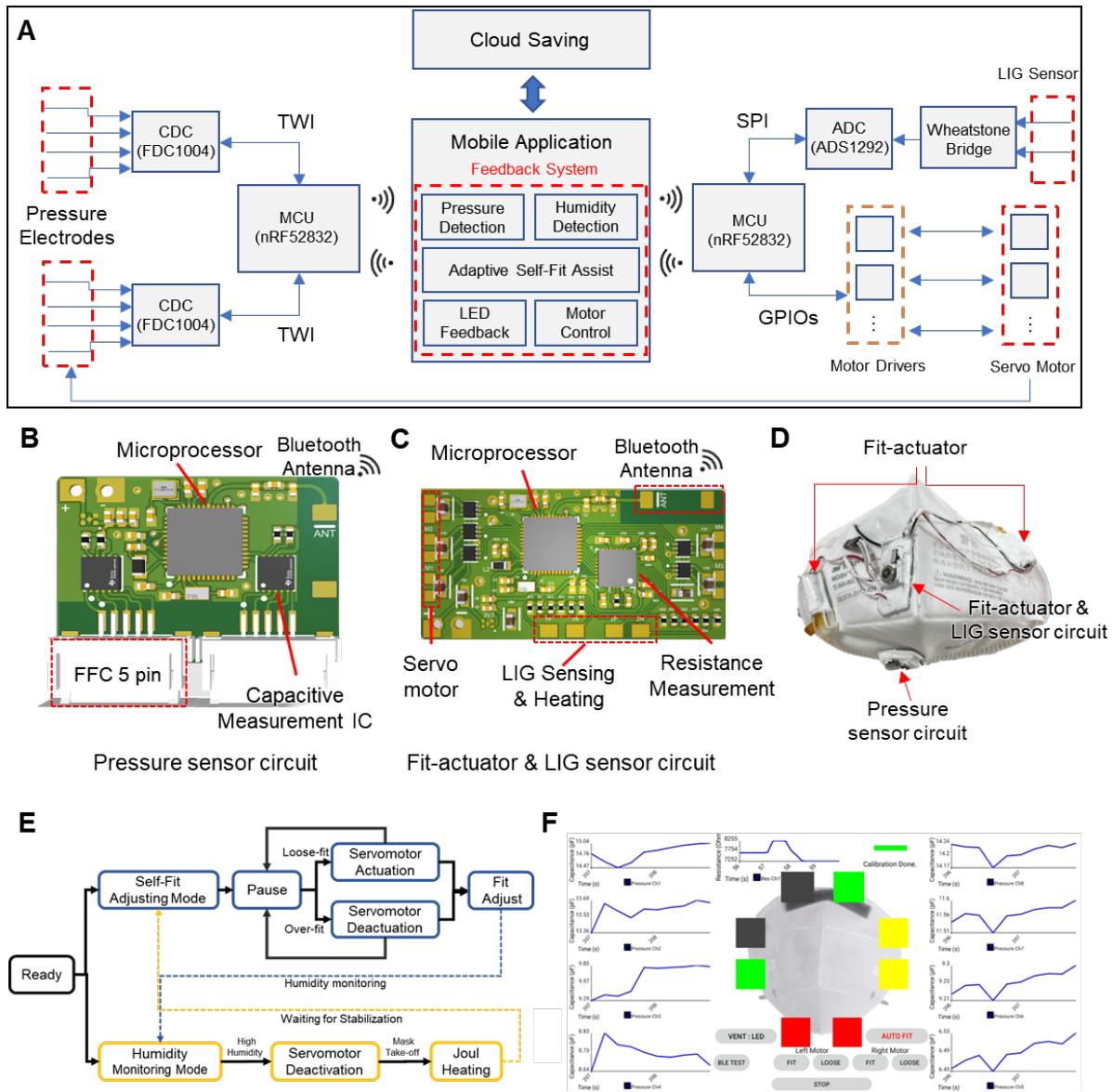

**Figure 4**

Wireless circuit development and device integration into the multifunctional smart respirator. A) Firmware operation logic and flow chart. B) Graphical illustration of the fPDB for wireless pressure sensing. C) Graphical illustration of the fPCB for wireless fit-actuating, LIG humidity sensing, and heating. D) Photograph of the smart respirator with embedded circuits and fit-actuators. E) Flowchart that describes the closed loop feedback control for fit adjusting and humidity sensing. F) Captured screenshot of the Android application developed in this work that shows the real-time pressure and humidity data along with a variety of functionalities.

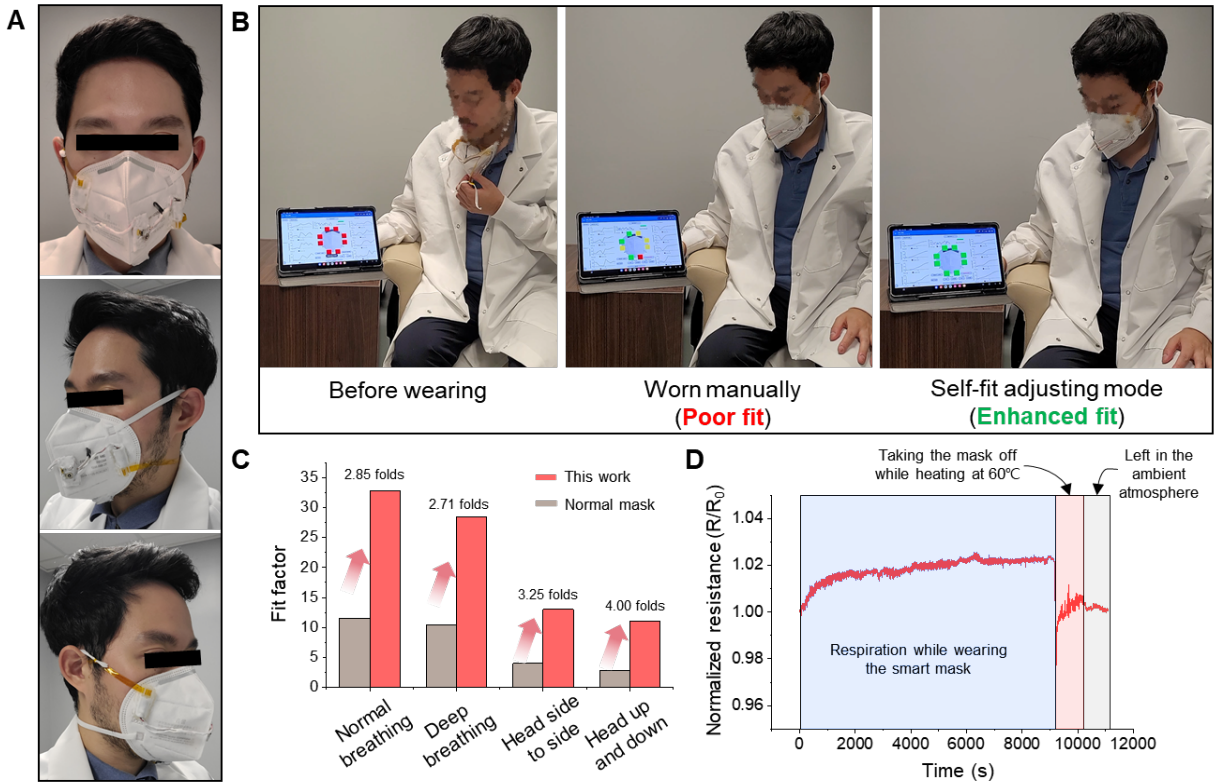

**Figure 5**

In vivo human subject test and performance examination. A) Front and side view of the human subject wearing the smart respirator. B) Demonstration of the self-fit adjusting mode on the human subject. D) Fit factor examination to compare with the commercially available respirator for four different scenarios. D) Humidity sensing plot within the respirator when the human subject wears the respirator and after taking it off while heating it at 60°C to facilitate the desorption process.